# Pressure-induced Superconductivity and Structure Phase Transition in SnAs-based Zintl Compound SrSn$_2$As$_2$

Weizheng Cao[1#], Juefei Wu[1#], Yongkai Li[2,3,4#], Cuiying Pei[1], Qi Wang[1,5], Yi Zhao[1], Changhua Li[1], Shihao Zhu[1], Mingxin Zhang[1], Lili Zhang[6], Yulin Chen[1,5,7], Zhiwei Wang[2,3,4*], Yugui Yao[2,3], and Yanpeng Qi[1,2,8*]

1. School of Physical Science and Technology, ShanghaiTech University, Shanghai 201210, China
2. Centre for Quantum Physics, Key Laboratory of Advanced Optoelectronic Quantum Architecture and Measurement (MOE), School of Physics, Beijing Institute of Technology, Beijing 100081, China
3. Beijing Key Lab of Nanophotonics and Ultrafine Optoelectronic Systems, Beijing Institute of Technology, Beijing 100081, China
4. Material Science Center, Yangtze Delta Region Academy of Beijing Institute of Technology, Jiaxing, 314011, China
5. ShanghaiTech Laboratory for Topological Physics, ShanghaiTech University, Shanghai 201210, China
6. Shanghai Synchrotron Radiation Facility, Shanghai Advanced Research Institute, Chinese Academy of Sciences, Shanghai 201203, China
7. Department of Physics, Clarendon Laboratory, University of Oxford, Parks Road, Oxford OX1 3PU, UK
8. Shanghai Key Laboratory of High-resolution Electron Microscopy, ShanghaiTech University, Shanghai 201210, China

\# These authors contributed to this work equally.
\* Correspondence should be addressed to Y.Q. (qiyp@shanghaitech.edu.cn) or Z.W. (zhiweiwang@bit.edu.cn)

## ABSTRACT

Layered SnAs-based Zintl compounds exhibit a distinctive electronic structure, igniting extensive research efforts in areas of superconductivity, topological insulators and quantum magnetism. In this paper, we systematically investigate the crystal structures and electronic properties of the Zintl compound SrSn$_2$As$_2$ under high-pressure. At approximately 20.8 GPa, pressure-induced superconductivity is observed in SrSn$_2$As$_2$ with a characteristic dome-like evolution of $T_c$. Theoretical calculations together with high pressure synchrotron X-ray diffraction and Raman spectroscopy have identified that SrSn$_2$As$_2$ undergoes a structural transformation from a trigonal to a monoclinic structure. Beyond 28.3 GPa, the superconducting transition temperature is suppressed due to a reduction of the density of state at the Fermi level. The discovery of pressure-induced superconductivity, accompanied by structural transitions in SrSn$_2$As$_2$, greatly expands the physical properties of layered SnAs-based compounds and provides a new ground states upon compression.

# INTRODUCTION

Zintl phases, a unique category of intermetallic compounds, were first introduced by E. Zintl in 1939, and have attracted considerable research attention[1]. Zintl compounds are characterized by their intricate chemical bonding and structural features. Their valence bonding modes can encompass ionic, metallic, and covalent interactions, reflecting their remarkable compositional and structural diversity[2, 3]. Zintl compounds are primarily categorized into layered, chain, cage, etc. These classifications often include $AB_2X_2$, $A_5B_2X_6$, $AX_3$, and $B_4X_3$, with A representing alkali, alkaline-earth or rare-earth metals, B representing transition metals and X representing metalloids[4-12]. With flexible structures, Zintl compounds possess a broad spectrum of physical properties, encompassing superconductivity, topological properties, magnetic order and thermoelectricity, etc[12-19].

In recent years, the layered SnAs-based Zintl compounds have attracted much attention[20-28]. Layered Zintl compound $NaSn_2As_2$ crystallizes in trigonal $R\bar{3}m$ structure, where $Na^+$ ions are separated by two honeycomb $[SnAs]^{2-}$ layers. The adjacent honeycomb layers interaction is *via* van der Waals (vdW) forces [25, 28]. At ambient pressure, $NaSn_2As_2$ showed bulk superconductivity with $T_c$ of 1.3 K[25, 28]. It should be noted that $NaSn_2As_2$ is a non-electron-balanced compound, containing $Sn^{2+}$ ions with lone pairs of electrons[22, 29]. In contrast to $NaSn_2As_2$, the isostructural compounds $EuSn_2As_2$ is electron-balanced. $EuSn_2As_2$ contains magnetic $Eu^{2+}$ ions, forming a peelable layered magnetic Zintl phase. A transition from paramagnetic (PM) to antiferromagnetic (AFM) phase in $EuSn_2As_2$ occurs around $T_N \sim 24$ K[30-32]. Below $T_N$, $EuSn_2As_2$ is ferromagnetic in the *ab* plane and antiferromagnetic between adjacent layers, forming an A-type AFM. A combination of first-principles calculations and angle-resolved photoemission spectroscopy (ARPES) experiments reveal that $EuSn_2As_2$ is a magnetic topological insulator (TI), characterizing by the absence of a detectable gap in the Dirac topological surface states (SSs). Besides, $EuSn_2As_2$ transforms from a strong TI with PM state to an axial insulator with AFM state below $T_N$[33]. $T_N$ shows a linear increase with pressure below 10 GPa, attributed to the enhanced interlayer magnetic exchange coupling among $Eu^{2+}$ ions[34]. Beyond ~14 GPa, $EuSn_2As_2$ experiences a two-step high-pressure structural transformation, giving rise to a novel monoclinic configuration. The bent Sn-Sn bonds become planar and form honeycomb Sn sheets, coinciding with the emergence of superconductivity around 4 K [35, 36].

$SrSn_2As_2$ is the sister compound of $EuSn_2As_2$, and $SrSn_2As_2$ remains relatively less explored. Theoretical calculations of the electronic structure propose that $SrSn_2As_2$ is a potential candidate for the novel three-dimensional Dirac semimetal due to its close proximity to the critical point[37]. The ARPES results present a band reversal feature near the Γ point, indicating that $SrSn_2As_2$ may be a new topological insulator[38]. Given that $EuSn_2As_2$ is superconducting accompanied with structure transition under high pressure, it is interesting to explore novel quantum phenomena in $SrSn_2As_2$ upon compression. Hence, we systematically investigate the structural and electronic properties of the SnAs-based Zintl compound $SrSn_2As_2$ under high-pressure. Interestingly, we observed the pressure-induced superconductivity in $SrSn_2As_2$, with a characteristic dome-shaped evolution of $T_c$. Our theoretical calculations reveal that $SrSn_2As_2$ undergoes a structural transformation from a trigonal to a monoclinic phase under high pressure, as evidenced by both X-ray diffraction (XRD) and Raman data. The electronic band structure of high-pressure phase and the evolution of $T_c$ are also discussed.

# EXPERIMENTAL DETAILS and CALCULATION METHODS

The single crystals of SrSn$_2$As$_2$ were grown by self-flux method. In order to obtain high-quality single crystals, pretreatment of starting materials (Sn, Alfa Aesar, 99.999% and As, Alfa Aesar, 99.99%) was performed to remove possible oxide layers on their surface by hydrogen reduction method and sublimation recrystallization method. High-purity starting materials of Sr, Sn, and As were loaded into an Al$_2$O$_3$ crucible with the atomic ratio of Sr: Sn: As = 1: 2: 2.2, and sealed into a quartz tube in a vacuum of $8 \times 10^{-4}$ Pa. The raw materials were reacted and homogenized at 1173 K for several hours, followed by cooling down to 773 K at a rate of 3 K/h. The crystalline phase of SrSn$_2$As$_2$ was checked by the X-ray diffraction (XRD, Cu $K\alpha$, $\lambda$ = 1.54184 Å). The chemical composition of SrSn$_2$As$_2$ is given by energy-dispersive x-ray spectra (EDX). Electrical transport properties were performed on a physical property measurement system (PPMS, Quantum Design).

Electrical transport measurements under higher pressures were performed in a nonmagnetic diamond anvil cell (DAC)[39-41]. A cubic BN/epoxy mixture layer was inserted between BeCu gaskets and electrical leads. Four platinum sheet electrodes were touched to the sample for resistance measurements with the van der Pauw method[40, 42, 43]. Pressure was determined by the ruby luminescence method[44]. High-pressure *in situ* Raman spectroscopy investigation was performed using a Raman spectrometer (Renishaw in-Via, UK) with a laser excitation wavelength of 532 nm and a low-wavenumber filter. A symmetric DAC with anvil culet sizes of 300 μm was used, with silicon oil as pressure transmitting medium (PTM). High-pressure *in situ* XRD measurements were performed at beamline BL15U of Shanghai Synchrotron Radiation Facility (X-ray wavelength $\lambda$ = 0.6199 Å). A symmetric DAC with anvil culet sizes of 200 μm and Re gaskets were used. Silicon oil was used as the PTM. The two-dimensional diffraction images were analyzed using the FIT2D softwar[45]. Rietveld refinements of crystal structure under various pressures were performed using the GSAS and the graphical user interface EXPGUI[46, 47].

We used the machine learning graph theory accelerated crystal structure search method (Magus) to explore the structures of SrSn$_2$As$_2$ under 30 GPa and 50 GPa[48, 49]. We performed the geometry optimization using the Vienna Ab-initio Simulation Package (VASP) based on the density functional theory [50, 51]. The exchange-correlation functional was treated by the generalized gradient approximation of Perdew, Burkey, and Ernzerhof [52]. The calculations used projector-augmented wave (PAW) approach to describe the core electrons and their effects on valence orbitals[53]. The plane-wave kinetic-energy cutoff was set to 600 eV, and the Brillouin zone was sampled by the Monkhorst-Pack scheme of $2\pi \times 0.03$ Å$^{-1}$. The convergence tolerance was $10^{-6}$ eV for total energy and 0.003 eV/Å for all forces. The electronic structure calculations used a denser k-mesh grid of $2\pi \times 0.02$ Å$^{-1}$. The phonon spectrum were calculated by the PHONOPY program package using the finite displacement method with the supercell $2 \times 2 \times 2$[54].

# RESULTS AND DISCUSION

Prior to high-pressure measurements, we first check the sample quality by single-crystal and powder XRD diffractions. The single-crystal XRD patterns on the flat surface of the sample shows sharp (00l) diffraction peaks [Figure 1(a)]. The calculated lattice parameter is $c$ = 26.66 Å, in agreement with the previous report[20]. The inset of Fig. 1(b) is the chemical compositional analysis results using EDX, illustrating a Sr:Sn:As atomic ratio of 20.45:38.97:40.59, which is consistent with the nominal composition. In addition, we further performed the powder XRD for structure characterizing. As shown in Fig. 1(b), all the Bragg peaks can be indexed into rhombohedral symmetry with the space group $R\bar{3}m$. The consistence between powder and single crystal XRD measurements guarantees the correct phase. The ambient crystal structure of SrSn$_2$As$_2$ is shown in Figure 1(c), which is identical to the configuration of EuSn$_2$As$_2$ and NaSn$_2$As$_2$. Then, we performed transport measurements at ambient pressure. Figure 1(d) shows the resistivity of SrSn$_2$As$_2$ as a function of temperature, showing typical metallic behavior with residual resistivity ratio (RRR) = 2.31.

Since NaSn$_2$As$_2$ showed superconductivity at ambient pressure and EuSn$_2$As$_2$ achieved superconductivity upon compression, it is natural to explore superconductivity in SrSn$_2$As$_2$ using high pressure technology. Hence, we investigated the effect of high-pressure on SrSn$_2$As$_2$ single crystals. Figure 2(a) shows the electrical resistivity $\rho(T)$ of SrSn$_2$As$_2$ at different pressures. Increasing pressure induces a continuous suppression of the overall magnitude of $\rho(T)$, which is typical behavior of metal under high pressure. At 14.9 GPa, the resistivity of SrSn$_2$As$_2$ drops abruptly at 1.8 K [Figure 2(b)]. As shown in Fig. 2(b), the resistivity dropped around 3 K and becomes more pronounced upon further compressing. Above 28.3 GPa, zero resistivity is observed at low temperatures, indicating a superconducting transition. The superconducting transition temperature $T_c$ (90% drop of the normal state resistivity) reaches 4.63 K at P = 28.3 GPa. As plotted in Fig. 2(c), $T_c$ decreases slowly beyond this pressure, and the superconductivity persists up to 53.5 GPa. The temperature dependence of transition width $\Delta T_c$ (10%–90% of the normal state resistance at $T_c$) is in Fig. 6. The transition width $\Delta T_c$ has a sharp decline from 2.07 K to 0.44 K in the pressure range from 22.9 GPa to 34.4 GPa. Transition width $\Delta T_c$ reflects the superconducting stated disturbance originating from the thermodynamic fluctuations, the applied magnetic field, the presence of secondary crystalline phases, the applied pressure, etc. [55], which needs further evidence to confirm its origin. The overall behavior of $T_c$ is a typical dome-like evolution under high pressure. Interestingly, a dome-like $T_c$ is observed during decompression, and the superconducting transition persists until recovery to 14.9 GPa (Fig. S1 in the Supplemental Material).

To gain insight into the superconducting transition, we applied an external magnetic field of 28.3 GPa and 48.01 GPa during the rise and fall of $T_c$ on SrSn$_2$As$_2$, respectively. Figure 2(d) and (e) demonstrate that the $T_c$ is continuously suppressed with increasing magnetic field and the superconducting transition could not be observed above 1.8 K at around 2.5 T. This confirms that the sharp drop of $\rho(T)$ around 4 K in SrSn$_2$As$_2$ originates from a pressure-induced superconducting transition. The upper critical field $\mu_0H_{c2}$ is determined from the 90% point on the resistivity transition curve，and the plot of temperature normalized $H_{c2}(T)$ is shown in Fig. 2(f). By fitting the data using the Ginzburg-Landau (GL) formula $\mu_0H_{c2}(T) = \mu_0H_{c2}(T)(1-t^2)/(1+t^2)$, where $t = T/T_c$ is the reduced temperature with zero-field superconducting $T_c$. The extrapolated upper critical fields $\mu_0H_{c2}(0)$ at 28.3 GPa and 48.1 GPa are 2.05 T and 2.41 T, which yields a Ginzburg-Landau coherence length $\xi_{GL}(0)$ of 12.68 nm and 11.69 nm, respectively.

The transition width $\Delta T_c$ drastic changed at around 30 GPa, and the slopes of d$H_{c2}$/d$T$ are notably different: − 0.53 and − 0.69 T/K for 28.3 and 48.1 GPa, respectively. Our results suggest that the nature of the superconducting state beyond 30 GPa may differ from that of the initial superconducting one. In order to identify the structural stability of SrSn$_2$As$_2$ under high pressure, we have performed high-pressure *in situ* synchrotron XRD and Raman spectroscopy measurements, as shown in Fig. 3. The XRD patterns of SrSn$_2$As$_2$ collected at different pressures are shown in Fig.

3(a). As the pressure increases, all diffraction peaks move to higher angles due to lattice contraction, and no structural phase transition is observed at pressures up to 29.8 GPa. Above 33.0 GPa, additional diffraction peaks appear, indicating a structural phase transition. Fig. 3(b) presents the Raman spectra of bulk SrSn$_2$As$_2$ under various pressures up to 55.5GPa. With increasing pressure, the interaction force between adjacent layers increases and all four phonon modes exhibit blue-shift, which is analogous to EuSn$_2$As$_2$[36]. The Raman signals of the $A_{1g}^2$ mode become significant, while $E_g^2$ mode decreases monotonically. An abrupt disappearance of Raman peaks for pressure beyond 33.6 GPa indicates the structural phase transition to a high-pressure phase. The evolution of the Raman spectra is consistent with our synchrotron XRD patterns and provides further evidence for pressure-induced structural phase transitions.

It should be emphasized that by only relying on the experimental data, the structural solution of high-pressure phases is not possible, because the XRD peaks are rather weak and broad. Hence, we performed the structure predictions at 30 GPa and 50 GPa. In each searching, structures were evaluated within 25 generations with 30 structures per generation, and the ambient stable structure $R\bar{3}m$ was treated as seed structure. We found one stable candidate $C2/m$ phase under high pressure, as shown in the inset of Fig. 4(a). The buckled Sn-Sn bonds become planar and form honeycomblike Sn sheets, meanwhile the SnAs layers further connect to each other *via* the As-As bonds across the Sr layers to form zigzag As chains between the Sn sheets. This three-dimensional monoclinic structure comprising honeycomblike Sn sheets and zigzag As chains resembles the situation in EuSn$_2$As$_2$ identified under high pressure [35]. As the enthalpy difference relative to $R\bar{3}m$ structure in Fig. 4(a), the enthalpy of $C2/m$ structure is below that of $R\bar{3}m$ above 22 GPa, suggesting that $C2/m$ structure is more energetically stable under high pressure. Then we calculated the phonon spectrum of $C2/m$ structure under high pressure, as plotted in Fig. 4(b) and (c). There are no imaginary frequencies in the phonon dispersion of $C2/m$ above 25 GPa, illustrating its dynamical stability. In summary, our theoretical and experimental results suggest that there is a structural phase transition from $R\bar{3}m$ phase to $C2/m$ phase under high pressure.

Next, we calculated the electronic structures of $C2/m$ structure under high pressure. As depicted in Fig. 5 (a) and (c), the valence bands and conduction bands cross the Fermi energy in the band structures of $C2/m$ phase, exhibiting typical metal characteristics. We can observe steep conduction bands crossing the Fermi energy, which is beneficial for superconductivity. The corresponding partial density of states (PDOS) are in Fig. 5 (b) and (d). The Sn atoms make main contribution at the Fermi energy, and the total density of states (DOS) at Fermi energy N(E$_f$) decrease from 5.2 states/formula at 25 GPa to 4.1 states/formula at 50 GPa, which agrees with the slow decreasing of $T_c$ from 28.3 GPa to 53.5 GPa.

Based on the aforementioned results, we can establish a $T_c$-$P$ phase diagram for SrSn$_2$As$_2$ as shown in Fig. 6. Superconductivity was observed at around 20.8 GPa. A domelike evolution is observed with a maximum $T_c$ of 4.63 K at 28.3 GPa for SrSn$_2$As$_2$. The high pressure *in-situ* synchrotron XRD and Raman spectroscopy reveal the evidence of structural transition around 28.3 GPa, which is in line with the theoretical predictions that the ambient $R\bar{3}m$ phase transforms to the high-pressure $C2/m$ phase. Combining the transition width results and theoretical calculations, the $T_c$-$P$ phase diagram reveals two distinct superconducting regions: SC-I $R\bar{3}m$ phase and SC-II $C2/m$ phase. In the SC-I region, $T_c$ increases with pressure with a broaden superconducting transition width. In the SC-I region between 25 and 60 GPa, $T_c$ is monotonically suppressed with external pressure. The suppression of $T_c$ in SrSn$_2$As$_2$ under pressure can be attributed to a decline in electronic density of states at the Fermi level.

## CONCLUSION

In summary, we have synthesized SrSn$_2$As$_2$ single crystal and explored the structure and electronic transport properties under high pressure. Our results demonstrate a pressure-induced superconductivity in SrSn$_2$As$_2$. The pressure dependent $T_c$ follows a dome-like evolution with a maximum $T_c$ value of 4.63 K at 28.3 GPa. Our theoretical calculations, together with high-pressure *in-situ* X-ray diffraction, Raman spectroscopy measurements, indicate that SrSn$_2$As$_2$ transforms from the ambient phase $R\bar{3}m$ to the monoclinic $C2/m$ phase above 25 GPa. Our research provides valuable insights into the understanding of the superconductivity in the layered SnAs-based family.

## ACKNOWLEDGMENT

This work was supported by the National Natural Science Foundation of China (Grant Nos. 52272265, U1932217, 11974246, 12004252), the National Key R&D Program of China (Grant No. 2018YFA0704300), and Shanghai Science and Technology Plan (Grant No. 21DZ2260400). Z.W.W. thanks the support from the National Key R&D Program of China (Grant Nos. 2020YFA0308800 and 2022YFA1403400), the Natural Science Foundation of China (Grant No. 92065109), the Beijing Natural Science Foundation (Grant Nos. Z210006 and Z190006). The authors thank the Analytical Instrumentation Center (# SPST-AIC10112914), SPST, ShanghaiTech University and the Analysis and Testing Center at Beijing Institute of Technology for assistance in facility support. The authors thank the staffs from BL15U1 at Shanghai Synchrotron Radiation Facility for assistance during data collection.

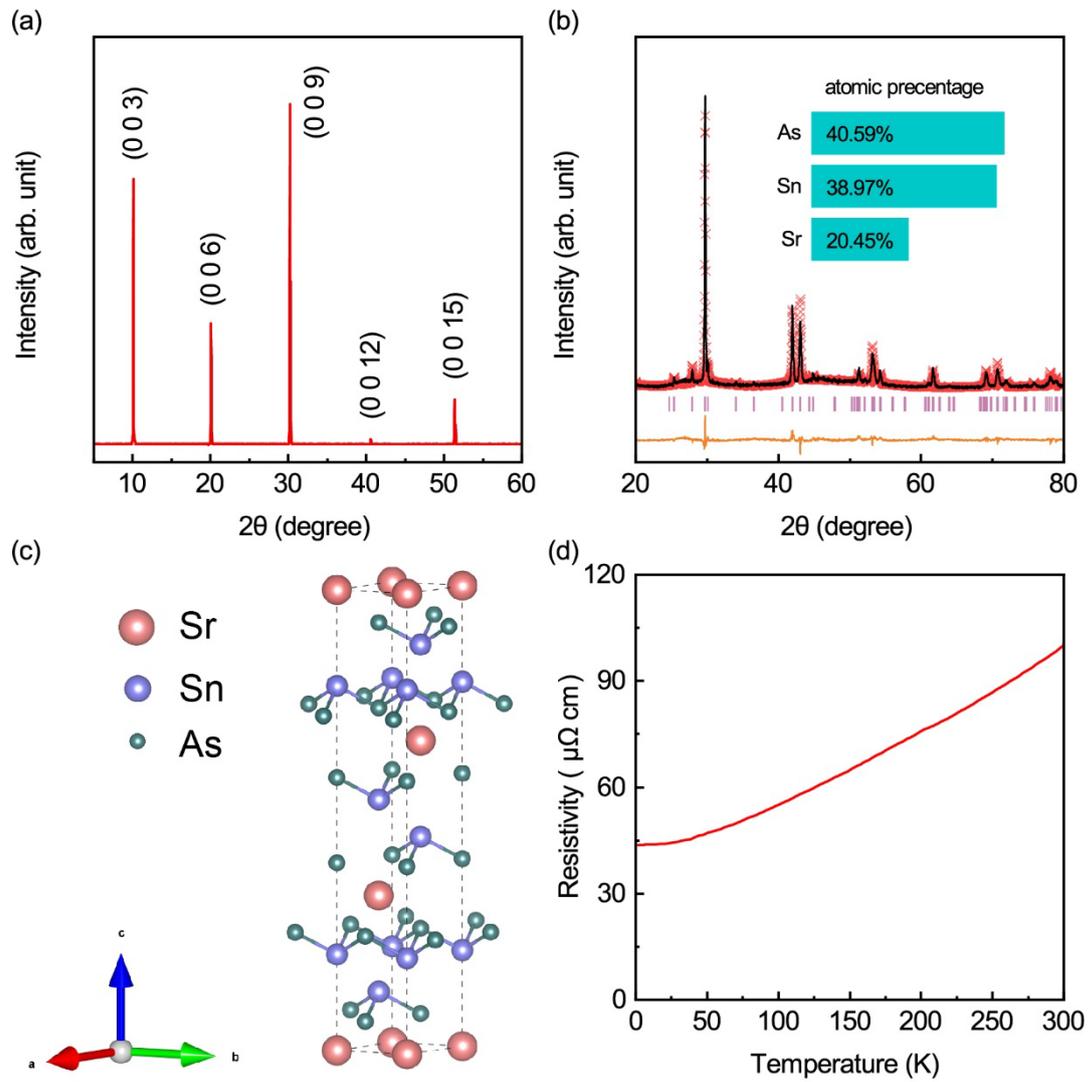

FIG. 1. (a) The crystal structure of SrSn$_2$As$_2$ with a space group $R\bar{3}m$. Pink, violet, and green balls represent Sr, Sn, and As atoms, respectively. (b) The X-ray diffraction peaks from the *ab* plane of SrSn$_2$As$_2$ single crystal. (c) Powder XRD pattern of SrSn$_2$As$_2$ at room temperature. Inset: the elemental content of SrSn$_2$As$_2$. (d) Resistivity dependence of temperature for SrSn$_2$As$_2$.

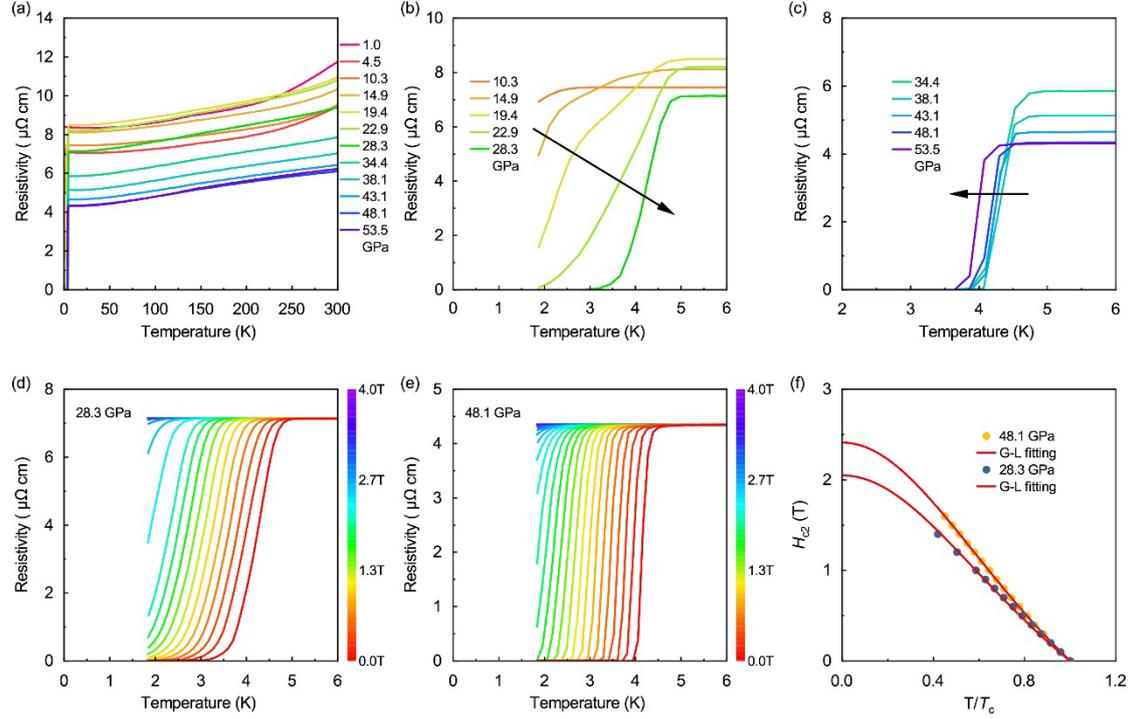

FIG. 2. (a) Pressure dependence of electrical resistivity of SrSn$_2$As$_2$ at a temperature range of 1.8–300 K. (b) and (c) Pressure induced superconductivity. The temperature-dependent resistivity under various pressures from 10.3 to 53.5 GPa. (d) and (e) Resistivity of SrSn$_2$As$_2$ as a function of temperature under different magnetic fields at pressure of 28.3 GPa and 48.1 GPa, respectively. (f) $H_{c2}$ as a function of normalized temperature at the pressure of SrSn$_2$As$_2$ at 28.3 GPa and 48.1 GPa, respectively. $T_c$ is determined as the 90% drop of the normal state resistivity. The solid lines represent the Ginzburg-Landau (G-L) fitting.

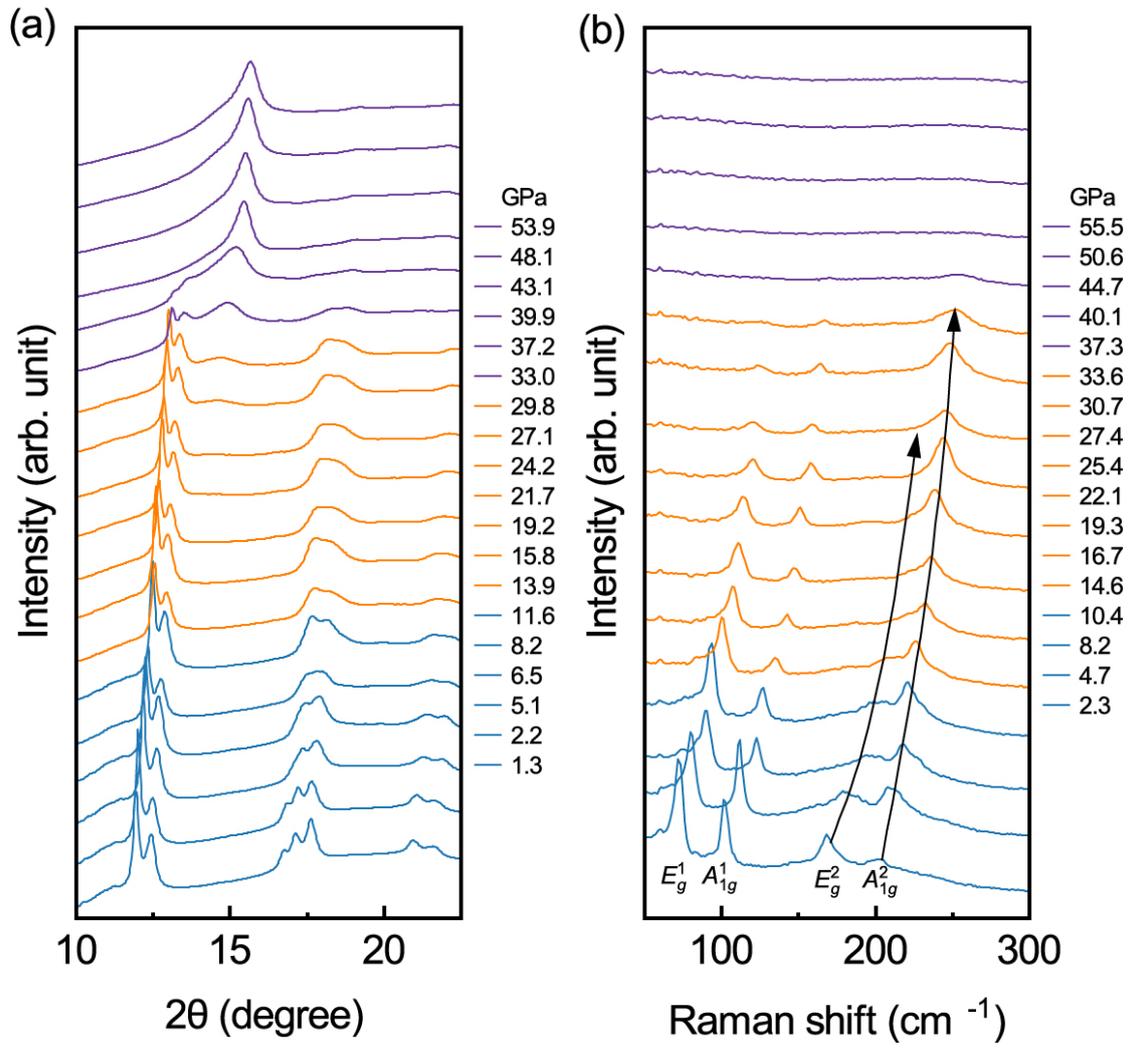

FIG. 3. (a) XRD patterns of SrSn$_2$As$_2$ under different pressures up to 53.9 GPa. (b) Raman spectra of SrSn$_2$As$_2$ under pressure at room temperature.

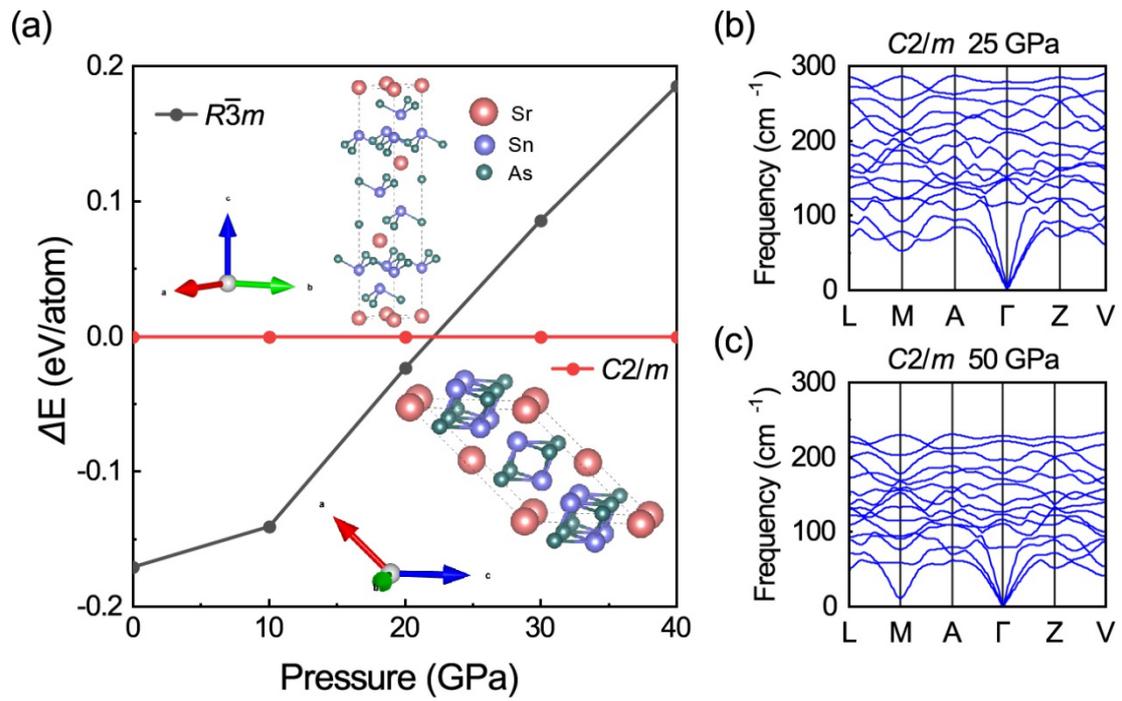

FIG. 4. (a) The enthalpy difference relative to $R\bar{3}m$ structure with in 40 GPa. The calculated phonon spectrum of the predicted $C2/m$ structure at (b) 25 GPa and (c) 50 GPa.

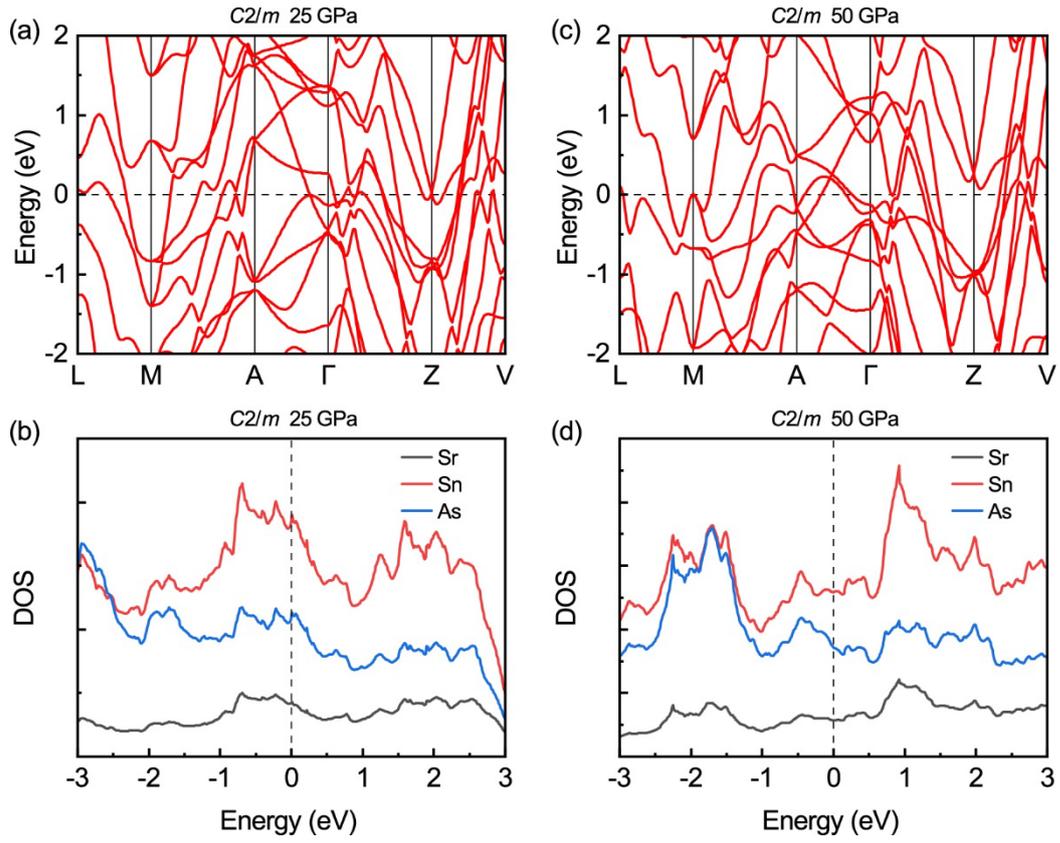

FIG. 5. The band structures of the predicted *C*2/*m* structure at (a) 25 GPa and (b) 50 GPa. The corresponding partial density of states (PDOS) at (c) 25 GPa and (d) 50 GPa, respectively. The Fermi energy

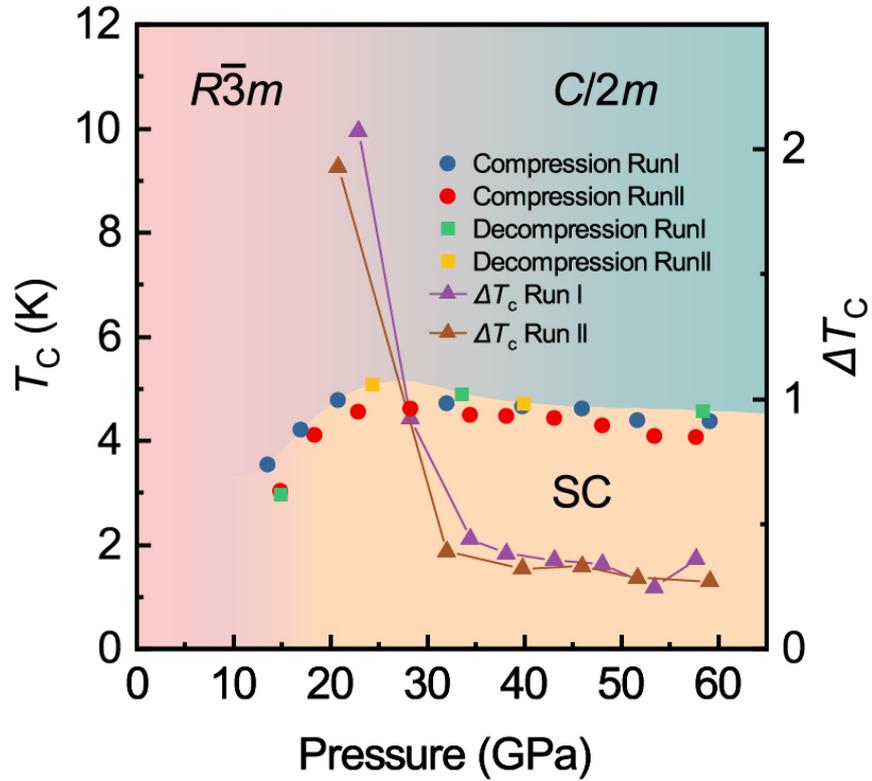

FIG.6. Phase diagram of SrSn$_2$As$_2$. Blue, red solid circles and green, yellow solid squares represent the $T_c$ in different compression and decompression experimental runs, respectively. The left vertical axis represents superconducting transition temperature. The beige background represents the superconducting transition temperature. A domelike evolution is observed with a maximum $T_c$ of 4.63 K at 28.3 GPa for SrSn$_2$As$_2$. The purple and brown triangular lines represent the pressure dependence of the superconducting transition width $\Delta T_c$. The right vertical axis represents superconducting transition width $\Delta T_c$. The pink and cyan backgrounds represent two different structures. SrSn$_2$As$_2$ undergoes a structural transformation from $R\bar{3}m$ to $C/2m$ at high pressure.